\documentstyle[12pt]{article}
\begin{document}
\begin{center}
{\large THE VARIATIONAL THEORY OF PERFECT FLUID WITH INTRINSIC HYPERMOMENTUM
IN SPACE-TIME WITH NONMETRICITY}\\
\vskip 0.4cm
BABOUROVA\enspace O.V. and FROLOV\enspace B.N. \\
\vskip 0.4cm
{\it Department of Mathematics, Moscow State Pedagogical University,\\
     Krasnoprudnaya 14, Moscow 107140, Russia;\\
     E-mail: baburova.physics@mpgu.msk.su,\enspace
frolovbn.physics@mpgu.msk.su}
\end{center}
\vskip 0.6cm
\begin{abstract}
{\small
\par
        The variational theory of the perfect  fluid with an intrinsic
hypermomentum is developed. The Lagrangian density of such fluid is stated
and the equations of motion of the fluid and the evolution equation of the
hypermomentum tensor are derived. The expressions of the matter currents of
the fluid (the metric stress-energy 4-form, the canonical energy-momentum
3-form and the hypermomentum 3-form) are obtained.}
\end{abstract}

%\newpage
\section{Introduction}
        The perfect  fluid  with an intrinsic hypermomentum as a new type of
matter was announced in \cite{Bab1}, \cite{Bab:thes}. The variational theory
of such fluid in a metric-affine space-time $(L_{4},g)$ \cite{He:nat} was
developed in \cite{Bab:Arg}-\cite{Bab:tr3}, in the paper \cite{Ob-Tr} the
exterior form language being used. The used variational method generalizes
the variational theory of the Weyssenhoff-Raabe perfect spin fluid based on
accounting the constraints in the Lagrangian density of the fluid with the
help of Lagrange multipliers, which has been developed  in case of a Riemann
-Cartan space-time in \cite{Tun1}-\cite{Bab:Boul} and in case of a metric
-affine space-time in \cite{Bab:thes}, \cite{Bab-Fr}. On the other
variational methods of the perfect spin fluid in a Riemann-Cartan space-time
see \cite{Min-Kar}, \cite{Kop}.
\par
        The theory of the perfect fluid with internal degrees of freedom
being developed, the additional internal degrees of freedom of a fluid
element are described by the four vectors $\bar{l}_{p}$ ($p = 1,2,3,4$),
called {\it directors}, adjoined with the each element of the fluid.
Three of the directors ($p = 1,2,3$) are space-like and the fourth ($p = 4$)
is time-like and is chosen to be equal to 4-velocity of the fluid element.
\par
        The distinctions in the various variational approaches consist in the
different properties the directors to be endowed. In \cite{Bab-Fr},
\cite{Bab:tr1}, \cite{Bab:tr2} the orthonormalization of the four directors is
maintained while a fluid element moving. In \cite{Ob-Tr} the three  space-like
directors are {\it elastic} in the sense that they can undergo
arbitrary deformations during the motion of the fluid and the orthogonality
of each of them to the 4-velocity is maintained. In \cite{Bab:tr3} none of
the orthogonality conditions of the four directors is maintained and all
directors are elastic.
\par
        The second distinction of the variational machinery consists in
using the generalized Frenkel condition $J^{\alpha}\!_{\beta} u^{\beta} =
J^{\alpha}\!_{\beta} u_{\alpha} = 0 $ \cite{Bab1}-\cite{Ob-Tr} or the Frenkel
condition in its classical form $S^{\alpha}\!_{\beta} u^{\beta} = 0 $
\cite{Bab:tr3}, where $J^{\alpha}\!_{\beta}$ and $S^{\alpha}\!_{\beta} =
J^{[\alpha}\!_{\beta ]}$ are the specific intrinsic hypermomentum tensor and
the specific spin tensor of a fluid element, respectively.
\par
        In this paper we use the exterior form variational method according
to Trautman \cite{Tr1} (see also \cite{Ob-Tr}, \cite{He:pr}). In our
approach it is essential that none of the orthogonality conditions of the
four directors is maintained during the motion of the fluid and the usual
Frenkel condition is valid in its usual form as in \cite{Bab:tr3}.

\section{The dynamical variables and constraints}
\setcounter{equation}{0}
         In the exterior form language the directors turn into the fields of
3-form $\tilde{l}_{q}$ and 1-form $\tilde{l}^{p}$ $(p = 1,2,3,4)$
representing the material frame and coframe, respectively, adjoined with
a fluid element, while the constraint
\begin{equation}
\tilde{l}^{p} \wedge \tilde{l}_{q} = \delta^{p}_{q} \tilde{\omega}
\label{eq:1}
\end{equation}
being fulfilled, where $\tilde{\omega}$ is the volume 4-form. We shall
consider the 1-form $\tilde{l}^{p}$ as an independent variable and
the 3-form $\tilde{l}_{q}$ as a function of $\tilde{l}^{p}$ by means of
(\ref{eq:1}). In the component representation one has
\begin{equation}
\tilde{l}^{p} = l^{p}_{\alpha}\tilde{\theta}^{\alpha}\; , \;\;\;\;\;
\tilde{l}_{q} = l_{q}^{\beta}\tilde{\omega}_{\beta}\; , \;\;\;\;\;
l^{p}_{\alpha}l_{p}^{\beta} = \delta_{\alpha}^{\beta} \; , \label{eq:2}
\end{equation}
where $\tilde{\theta}^{\alpha}$ is a basis 1-form and
$\tilde{\omega}_{\beta}$  is a 3-form defined as \cite{Tr1}
\begin{equation}
\tilde{\omega}_{\beta} = *\! \bar{e}_{\beta}\; , \;\;\;\;\;\;\;\;\;\;
\tilde{\theta}^{\alpha} \wedge \tilde{\omega}_{\beta} = \delta^{\alpha}
_{\beta} \tilde{\omega} \; . \label{eq:3}
\end{equation}
Here $*$ is the Hodge dual operator and $\bar{e}_{\beta}$ is a basis vector,
a coordinate system being nonholonomic in general.
\par
        Each fluid element possesses a 4-velocity vector $\bar{u}$ which is
corresponded to a velocity 1-form $\tilde{u} = \breve{g}(\bar{u},\cdots)$ and
a flow 3-form $*\bar{u}$ \cite{Tr2} with
\begin{equation}
*\!\bar{u} \wedge \tilde{u} = c^{2}\tilde{\omega}\;, \label{eq:4}
\end{equation}
that means the usual condition $\breve{g}(\bar{u},\bar{u}) = - c^{2}$.
\par
        The director field and the 4-velocity field are compatible in the
sense that the director $\bar{l}_{4}$ is oriented along the velocity:
$\bar{l}_{4} = c^{-1}\bar{u}$. This condition yields the constraint
\begin{equation}
*\!\bar{u} \wedge \tilde{l}^{p} = - c\delta^{p}_{4}\tilde{\omega}\;.
\label{eq:5}
\end{equation}
\par
        A fluid element moving, the mass and entropy conservation laws are
fulfilled,
\begin{eqnarray}
d(\mu *\!\bar{u}) = 0 \; , \label{eq:6} \\
d(\mu s *\!\bar{u}) = 0 \; , \label{eq:7}
\end{eqnarray}
where $\mu$ and $s$ are the mass density  and the specific entropy of the
fluid in the rest frame of reference, respectively.
\par
        An element of the fluid with intrinsic  hypermomentum possesses
the additional kinetic energy 4-form
\begin{equation}
E =  \frac{1}{2} \mu J^{p}\!_{q}\Omega^{q}\!_{p}\tilde{\omega}\; ,
\label{eq:8}
\end{equation}
where $J^{p}\!_{q}$ is the specific intrinsic hypermomentum tensor
representing the new dynamical quantity which generalizes the spin density
of the Weyssenhoff fluid. The quantity $\Omega^{q}\!_{p}$ is the measure of
ability of a fluid element to perform the intrinsic motion and
generalizes the fluid element angular velocity of the Weyssenhoff spin fluid
theory. It has the form
\begin{equation}
\Omega^{q}\!_{p} \tilde{\omega} = *\!\bar{u} \wedge l^{q}_{\alpha}
{\cal D} l^{\alpha}_{p} \; , \label{eq:9}
\end{equation}
where ${\cal D}$ means the exterior covariant derivative,
\begin{equation}
{\cal D} l^{\alpha}_{p} = d l^{\alpha}_{p} + \tilde{\Gamma}^{\alpha}\!
_{\beta} l^{\beta}_{p} \; . \label{eq:10}
\end{equation}
\par
        The specific intrinsic hypermomentum tensor $J^{p}\!_{q}$ can be
decomposed into irreducible parts
\begin{equation}
J^{p}\!_{q} = S^{p}\!_{q} + \overline{J}^{p}\!_{q} + \frac{1}{4}
\delta^{p}_{q}J\;, \qquad S^{p}\!_{q}:= J^{[p}\!_{q]}\;, \qquad
\overline{J}^{p}\!_{p} = 0\;. \label{eq:11}
\end{equation}
Here $S^{p}\!_{q}$ is the specific spin tensor of a fluid element and
$J$ is the specific dilaton charge of a fluid element, respectively.
The former one obeys the Frenkel condition $S^{p}\!_{q}u^{\alpha}l^{q}
_{\alpha} = 0$ which can be represented in the form
\begin{equation}
J^{[p}\!_{q]} *\!\bar{u} \wedge \tilde{l}^{q} = 0 \; . \label{eq:12}
\end{equation}
\par
        The perfect fluid Lagrangian density 4-form should be chosen as the
remainder after subtraction the internal energy density of the fluid
$\varepsilon$ from the kinetic energy (\ref{eq:8}) with regard to the
constraints (\ref{eq:4})-(\ref{eq:7}), (\ref{eq:9}), (\ref{eq:12}) which
should be introduced into the Lagrangian density by means of the Lagrange
multipliers $\lambda$, $\nu_{p}$, $\varphi$, $\tau$, $\kappa^{p}\!_{q}$,
$\chi_{p}$, respectively.
\par
        The internal energy density of the fluid $\varepsilon$ depends on
the extensive (additive) thermodynamic parameters $\mu$, $s$, $J^{p}\!_{q}$
and obeys to the first thermodynamic principle
\begin{equation}
d\varepsilon(\mu, s, J^{p}\!_{q}) = \frac{\varepsilon + p}{\mu} d\mu +
\mu T ds + \frac{\partial \varepsilon}{\partial J^{p}\!_{q}} dJ^{p}\!_{q}
\; . \label{eq:13}
\end{equation}
\par
        We need the following variation of the dependent variables
which can be derived as a result of the resolution of the constrants
(\ref{eq:1}), (\ref{eq:2}) with the help of the relations (\ref{eq:3}),
\begin{eqnarray}
&&\tilde{\omega} \delta l^{p}_{\alpha} = l^{p}_{\beta} \tilde{\omega}_{\alpha}
\wedge \delta \tilde{\theta}^{\beta} - \tilde{\omega}_{\alpha} \wedge \delta
\tilde{l}^{p} \; , \label{eq:14} \\
&&\tilde{\omega} \delta l_{p}^{\alpha} = - \tilde{l}_{p} \wedge \delta
\tilde{\theta}^{\alpha} + \tilde{l}_{p} l_{q}^{\alpha} \wedge \delta
\tilde{l}^{q} \; . \label{eq:15}
\end{eqnarray}
As a result of the relation $*\!\bar{u} \wedge \tilde{\theta}^{\alpha} =
- u^{\alpha} \tilde{\omega}$ one also has
\begin{eqnarray}
&&\tilde{\omega} \delta u^{\alpha} = - \delta (*\!\bar{u}) \wedge
\tilde{\theta}
^{\alpha} - *\!\bar{u} \wedge \delta \tilde{\theta}^{\alpha}  - u^{\alpha}
\delta \tilde{\omega}\; , \label{eq:16} \\
&&\delta \tilde{\omega} = \tilde{\omega}\frac{1}{2} g^{\sigma\rho}\delta
g_{\sigma\rho} + \delta \tilde{\theta}^{\sigma} \wedge \tilde{\omega}
_{\sigma} \; . \label{eq:17}
\end{eqnarray}

\section{The Lagrangian density and the equations of motion of the fluid}
\setcounter{equation}{0}
        As a result of the previous section the Lagrangian density 4-form of
the perfect  fluid  with an intrinsic hypermomentum has the form
\begin{eqnarray}
\tilde{L}_{m} = L_{m} \tilde{\omega} =
- \varepsilon (\mu, s, J^{p}\!_{q}) \tilde{\omega} +
\frac{1}{2}\mu J^{p}\!_{q} \Omega^{q}\!_{p} \tilde{\omega} + \mu *\!\bar{u}
\wedge d\varphi + \mu\tau *\! \bar{u} \wedge ds \nonumber\\
+ \mu \lambda (*\!\bar{u} \wedge \tilde{u} - c^{2} \tilde{\omega}) +
\mu \nu_{p}(*\!\bar{u} \wedge \tilde{l}^{p} + c\delta^{p}_{(4)} \tilde
{\omega}) + \mu \chi_{p} J^{[p}\!_{q]} *\!\bar{u} \wedge \tilde{l}^{q}
\nonumber\\
+ \mu \kappa^{p}\!_{q}\, (*\!\bar{u} \wedge l^{q}_{\alpha}
{\cal D} l^{\alpha}_{p} - \Omega^{q}\!_{p} \tilde{\omega}) \;. \label{eq:20}
\end{eqnarray}
\par
        The fluid motion equations and the evolution equations of  the
hypermomentum tensor are derived by the variation of (\ref{eq:20}) with
respect to the independent variables $\mu$, $s$, $J^{p}\!_{q}$,
$\Omega^{q}\!_{p}$, $*\!\bar{u}$, $\tilde{l}^{p}$ and the Lagrange multipliers.
As a result
of such variational machinery one gets the constraints (\ref{eq:4})-
(\ref{eq:7}), (\ref{eq:9}), (\ref{eq:12}) and the following variational
equations,
\begin{eqnarray}
\delta \mu : &&\qquad \qquad  -(\varepsilon + p) \omega + \frac{1}{2}\mu
J^{p}\!_{q}\Omega^{p}\!_{q} \omega + \mu *\!\bar{u} \wedge d\varphi = 0\; ,
\label{eq:21}\\
\delta s : &&\qquad \qquad  T \tilde{\omega} + *\!\bar{u} \wedge
d\tau = 0 \; , \label{eq:22}\\
\delta J^{p}\!_{q} : &&\qquad\qquad \frac{\partial \varepsilon}
{\partial J^{p}\!_{q}} = \frac{1}{2} \mu \Omega^{q}\!_{p} - \mu c \chi_{[p}
\delta^{q]}_{(4)} \; ,  \label{eq:23}\\
\delta\Omega^{q}\!_{p} : &&\qquad\qquad \kappa^{p}\!_{q} =
\frac{1}{2} J^{p}\!_{q} \; , \label{eq:24}\\
\delta *\!\bar{u} : &&\qquad\qquad d\varphi + \tau ds + 2\lambda \tilde{u} +
\nu_{p}
\tilde{l}^{p} + \chi_{p} J^{[p}\!_{q]} \tilde{l}^{q} + \kappa^{p}\!_{\alpha}
{\cal D} l^{\alpha}_{p} = 0\; , \label{eq:25} \\
\delta \tilde{l}^{q} : &&\qquad\qquad \nu_{q} *\!\bar{u} + \chi_{p}
J^{[p}\!_{q]}
*\!\bar{u} + \frac{1}{2} \dot{J}^{\alpha}\!_{\beta} l^{\beta}_{q}
\tilde{\omega}_{\alpha} = 0\; . \label{eq:26}
\end{eqnarray}
In the equation (\ref{eq:26}) the notation
\begin{equation}
\dot{J}^{\alpha}\!_{\beta} = *\!(*\!\bar{u} \wedge {\cal D}J^{\alpha}\!
_{\beta})  \label{eq:27}
\end{equation}
was introduced.
\par
        Let us derive some consequenses of these equations. Multiplying the
equation (\ref{eq:25}) by $*\!\bar{u}$ from the left externally and using
(\ref{eq:24}) and (\ref{eq:21}) one derives the expression for the Lagrange
multiplier $\lambda$:
\begin{equation}
2 \mu c^{2} \lambda = -(\varepsilon + p) + \mu c \nu_{(4)} \; . \label{eq:28}
\end{equation}
Multiplying the equation (\ref{eq:26}) by $\tilde{u}$ from the right
externally one gets:
\begin{equation}
\nu_{q} + \chi_{p} S^{p}\!_{q} = \frac{1}{2c^{2}} \dot{J}^{\alpha}\!_{\beta}
l^{\beta}_{q} u_{\alpha} \; . \label{eq:29}
\end{equation}
This equation with regard of the Frenkel condition (\ref{eq:12}) has the
consequense
\begin{equation}
c\nu_{(4)} = \frac{1}{2c^{2}} \dot{J}^{\alpha}\!_{\beta} u_{\alpha}
u^{\beta} \; . \label{eq:30}
\end{equation}
\par
        The Lagrange maltiplier $\chi_{p}$ can be found as a consequense of
the correspondence principle of the theory under cosideration to the
Weyssenhoff spin fluid theory. Namely, the quantity canonically conjugated to
the spin tensor should be the spatial angular velocity of the directors,
\begin{equation}
\frac{\partial \varepsilon}{S^{p}\!_{q}} = \frac{\partial \varepsilon}
{J^{[p}\!_{q]}} = \frac{1}{2} \mu \Pi^{[q}_{r} \Pi^{s}_{p]} \Omega^{r}\!_
{s} \; , \qquad \Pi^{\alpha}_{\gamma} = \delta^{\alpha}_{\gamma}
+ \frac{1}{c^{2}} u^{\alpha} u_{\gamma} \; . \label{eq:341}
\end{equation}
Comparing (\ref{eq:341}) with (\ref{eq:23}) one can derive
\begin{equation}
\chi_{q} \Pi^{q}_{p} = \frac{1}{c^2} \Omega_{[pq]} u^{q} \; . \label{eq:342}
\end{equation}
Obviously that the Lagragian density (\ref{eq:20}) determines only the
spatial part of the Lagrange miltiplier $\chi_{p}$ and therfore the condition
$c\chi_{4} = \chi_{p} u^{p} = 0$ can be imposed without loss of generality.
Thus one has
\begin{equation}
\chi_{p} = \frac{1}{c^2} \Omega_{[pq]} u^{q} \; . \label{eq:343}
\end{equation}
\par
        As a consequense of the fluid motion equations it is easy to verify
that the Lagrangian density 4-form (\ref{eq:20}) is proportional to the
hydrodynamic fluid pressure,
\begin{equation}
\tilde{L}_{m} = p \tilde{\omega} \; . \label{eq:33}
\end{equation}
\par
        Substituting (\ref{eq:29}) into (\ref{eq:26}) one finds the  evolution
equation of the hypermomentum tensor,
\begin{equation}
\Pi^{\alpha}_{\gamma} \dot{J}^{\gamma}\!_{\beta} = 0 \;.  \label{eq:31}
\end{equation}
This equation has the consequense
\begin{equation}
\dot {J} + \frac{1}{c^{2}} \dot {J}^{\alpha}\!_{\beta} u_{\alpha} u^{\beta}
=  0\; , \qquad J = J^{\alpha}\!_{\alpha} \; . \label{eq:32}
\end{equation}

\section{The energy-momentum tensor of the perfect  fluid  with an
intrinsic hypermomentum}
\setcounter{equation}{0}
        The matter Lagrangian density makes possible to derive the external
currents of a matter field which are the sources of the gravitational field.
In case of the perfect fluid with an intrinsic hypermomentum one has as the
matter currents: the metric stress-energy 4-form $\tilde{T}^{\sigma\rho}$,
the canonical energy-momentum 3-form $\tilde{t}_{\sigma}$ and the
hypermomentum 3-form $\tilde{J}^{\alpha}\!_{\beta}$, which are determined as
variational derivatives \cite{He:pr}.
\par
        By virtue of the explicit form the Lagrangian density (\ref{eq:20})
the metric stress-energy 4-form reads
\begin{equation}
\tilde{T}^{\sigma\rho} = 2\frac{\delta\tilde{L}_{m}}{\delta g_{\sigma\rho}} =
T^{\sigma\rho} \tilde{\omega} \; , \qquad  T^{\sigma\rho} = g^{\sigma\rho}
p + \frac{1}{c^{2}}(\varepsilon + p + \frac{1}{2} \mu \dot{J}) u^{\sigma}
u^{\rho} \; . \label{eq:35}
\end{equation}
\par
        With regard to (\ref{eq:28}) and (\ref{eq:30}) the variational
derivative of (\ref{eq:20}) with respect to $\tilde{\theta}^{\sigma}$ yields
\begin{equation}
\tilde{t}_{\sigma} = \frac{\delta\tilde{L}_{m}}{\delta\tilde{\theta}^{\sigma}}
 =  p\tilde{\omega}_{\sigma} + \frac{1}{c^{2}}(\varepsilon + p + \frac{1}{2}
 \dot{J})u_{\sigma}*\!\bar{u} + \frac{1}{2}\mu \dot{J}^{\alpha}\!_{\sigma}
 \tilde{\omega}_{\alpha} \; . \label{eq:36}
\end{equation}
On the basis of the evolution equation of the hypermomentum tensor
(\ref{eq:31}) the expression of the canonical energy-momentum 3-form takes
the form \cite{GR14}
\begin{eqnarray}
\tilde{t}_{\sigma} =  p\tilde{\omega}_{\sigma} + \frac{1}{c^{2}}(\varepsilon
+ p + \frac{1}{2} \dot{J})u_{\sigma}*\!\bar{u} - \frac{1}{2c^{2}}\mu
\dot{J}^{\alpha}\!_{\sigma} u_{\alpha}*\!\bar{u} \nonumber \\
= p\tilde{\omega}_{\sigma} + \frac{1}{c^{2}} (\varepsilon + p) u_{\sigma}
*\!\bar{u} - \frac{1}{2c^{2}} \mu \dot{J}^{\beta}\!_{\alpha}
\Pi^{\alpha}_{\sigma} u_{\beta} *\!\bar{u} \; . \label{eq:37}
\end{eqnarray}
\par
        For the hypermomentum 3-form expression one finds
\begin{equation}
\tilde{J}^{\alpha}\!_{\beta} = - \frac{\delta\tilde{L}_{m}}{\delta\tilde
{\Gamma}^{\beta}\!_{\alpha}} = \frac{1}{2} \mu J^{\alpha}\!_{\beta}*\!\bar{u}
\; . \label{eq:38}
\end{equation}
\par
        The expressions of the metric stress-energy 4-form (\ref{eq:35}),
the canonical energy-momentum 3-form (\ref{eq:37}) and the hypermomentum
3-form (\ref{eq:38}) are compatible in the sense that they satisfy to the
Noether identity
\begin{equation}
T^{\sigma}\!_{\rho} \tilde{\omega} + \frac{1}{2} \mu *\!\bar{u} \wedge
{\cal D} J^{\sigma}\!_{\rho} = \tilde{\theta}^{\sigma} \wedge \tilde{t}_
{\rho} \; , \label{eq:39}
\end{equation}
that corresponds to the $GL(n,R)$-invariance of the Lagrangian density
(\ref{eq:20}) \cite{He:pr}.

\section{Conclusion}
        The essential feature of the constructed variational theory of the
perfect fluid with an intrinsic hypermomentum is the assumption that the
frame  realized by the all four directors is elastic and can be deformed
during the motion of the fluid element according to nonmetricity of the
space-time. Therefore the Lagrangian density (\ref{eq:20}) does not contain
the term maintaining the orthogonality of the directors. The second
distinction from the other variational approaches is the using of the Frenkel
condition in its classical form (\ref{eq:12}).
\par
        As a result we have obtained the new expression for the energy-
momentum tensor of the fluid which is the source of the gravitational field
in the metric-affine space-time. It should be important to investigate the
consequences of the employing of this energy-momentum tensor to cosmological
and astrophysical problems. For example, it is interesting to clarify whether
the corresponding field equations have the regular solution with the upper
limit for $\epsilon$ (the limiting energy density of the fluid). These
questions are under consideration now.

\section{Acknowledgement}
        This paper is partly supported by the scientific programm
"Univesitety Rossii".

\end{document}